\documentclass[preprint,12pt]{elsarticle}

\usepackage[utf8]{inputenc}
\UseRawInputEncoding
\usepackage{amsmath,amssymb}
\usepackage{makeidx}
\usepackage{graphicx}
\usepackage{bm,times}
\usepackage{multirow}
\usepackage{slashbox}
\usepackage{txfonts}
\usepackage{algorithm}
\usepackage{algorithmic}
\usepackage{colortbl}
\begin{document}

\begin{frontmatter}



\title{Bayesian data combination model \\ with Gaussian process latent variable model \\ 
for mixed observed variables under NMAR missingness}


\author[label1]{Masaki Mitsuhiro}
\author[label2,label3]{Takahiro Hoshino}

\address[label1]{Nikkei Research Inc.}
\address[label2]{Keio University}
\address[label3]{RIKEN Center for Advanced Intelligence Project}

\begin{abstract}

In the analysis of observational data in social sciences and businesses, it is difficult to obtain a ``(quasi) single-source dataset" in which the variables of interest are simultaneously observed. Instead, multiple-source datasets are typically acquired for different individuals or units.
Various methods have been proposed to investigate the relationship between the variables in each dataset, e.g., matching and latent variable modeling. It is necessary to utilize these datasets as a single-source dataset with missing variables.
Existing methods assume that the datasets to be integrated are acquired from the same population or that the sampling depends on covariates. This assumption is referred to as missing at random (MAR) in terms of missingness.
However, as will been shown in application studies, it is likely that this assumption does not hold in actual data analysis and the results obtained may be biased.
We propose a data fusion method that does not assume that datasets are homogenous. We use a Gaussian process latent variable model for non-MAR missing data. This model assumes that the variables of concern and the probability of being missing depend on latent variables.
A simulation study and real-world data analysis show that the proposed method with a missing-data mechanism and the latent Gaussian process yields valid estimates, whereas an existing method provides severely biased estimates.
This is the first study in which non-random assignment to datasets is considered and resolved under resonable assumptions in data fusion problem.

\end{abstract}

\begin{keyword}
Missing data \sep Bayesian inference \sep Shared-parameter model \sep Conditional independence assumption \sep Statistical data fusion \sep Gaussian Process latent variable model


\end{keyword}

\end{frontmatter}


\section{Introduction}
\begin{figure}[tbhp]
  \begin{center}
    \includegraphics[scale=0.34]{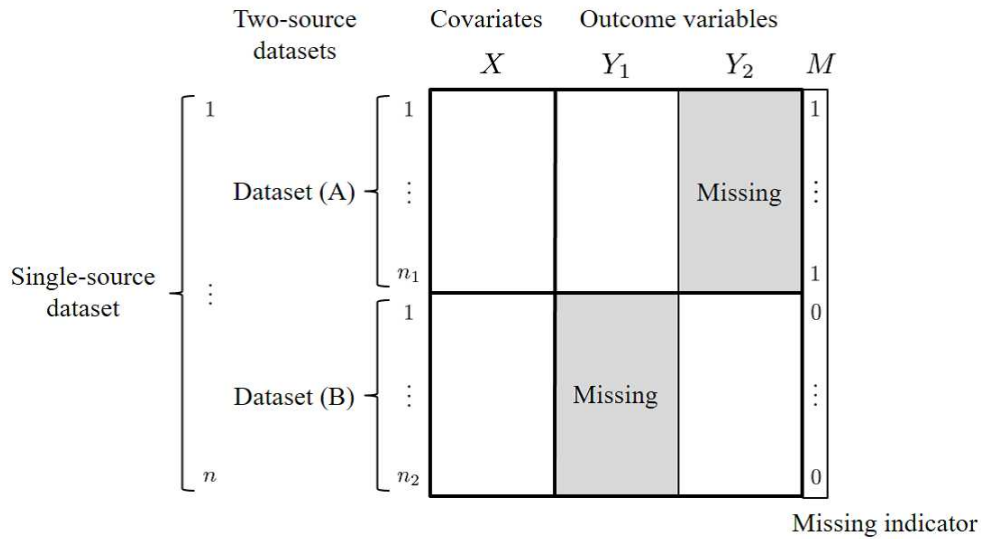}
  \end{center}
  \caption{Typical situation showing two-source datasets obtained in fields such as marketing}
  \label{datadef}
\end{figure}

It is necessary to investigate the relationship between simultaneously obtained variables to understand the causal mechanism behind the phenomena in various fields such as marketing and economics.
However, it is frequently difficult to obtain a single-source dataset. Instead, there are ``multiple-source datasets," in which different sets of variables are observed for different datasets.
Various methods have been proposed to infer the relationship between the variables of concern using these datasets.
For example, in the marketing field, consider that the purchase history ($Y_1$ in Figure \ref{datadef}) and demographic information ($X$) are observed for $n_1$ customers from the ID-POS transactions for a specific retailer (dataset (A)). In addition, the total spending in this category ($Y_2$) and demographic information ($X$) are observed for $n_2$ consumers from marketing research data (dataset (B)).
Then, the total spending $Y_2$ cannot be obtained for customers in ID-POS. Similarly, $Y_1$ is missing for consumers in the marketing research dataset.
We can obtain a ``(quasi) single-source dataset" (Assael and Poltrack, 1996) by predicting and imputing missing values. This dataset is used for marketing strategies such as segmentation, recommendation, and customer management.　
In actual data analysis, marketing managers are typically interested in the relationship between $Y_1$ and $Y_2$ or a function of these variables, such as the share of wallet (spending at a retailer divided by the total spending $Y_1/Y_2$). However, invalid imputation methods will yield a fallacy of inference.

This situation is well known as statistical data fusion (Kamakura and Wedel, 1997) in marketing and statistical data combination (Ridder and Moitt, 2007) in economics, in which multiple-source datasets obtained from different units are considered as a single-source dataset, as shown in Figure  \ref{datadef}.
The data combination problem is not limited to marketing. It is common in various fields such as finance, economy, medicine, and policy planning. Several methods have been proposed to solve this problem.

The most common strategy is statistical matching, which identifies pairs of units or individuals to match on the basis of the variables that are common to both datasets ($X$).
A pair is treated as the same unit, and the missing $Y_1$ for one pair is replaced with the observed $Y_1$ for the other pair.
Widely used statistical matching methods include Mahalanobis matching (MM) (Rubin, 1980) and propensity score matching (Rosenbaum and Rubin, 1983), which uses the balancing weights of covariates. 
However, the prediction accuracy of statistical matching is not high. Recently new matching techniques are developed in estimating causal effects. Endres and Augustin (2016) used probabilistic graphical models for statistical matching and Kallus (2020) proposed a generalized optimal matching, but in data combination problem, the target of infeference are not the causal estimands but the correlations or joint distributions, which makes applications of recent matching methods to data combination difficult.
Thus, various methods have been proposed to improve the accuracy of predicting unobserved variables, such as a regression model based on the Bayesian approach (Gilula et al., 2006), a semiparametric Bayes model  based on the prior distribution of a directory process (Hoshino, 2013), and a nonparametric model that uses the odds ratio (Qian and Xie, 2014).

Existing methods assume that the datasets to be integrated are obtained from the same population or that sampling depends on covariates. This assumption is known as missing at random (MAR) (Rubin, 1976) in terms of missing data and missingness.
However, it is quite likely that this assumption does not hold in actual data analysis.
In the abovementioned example, the customers in dataset (A) can use their loyalty cards to obtain rewards from a retailer. Thus, these customers will purchase more from the retailer compared to the consumers in dataset (B), who are sampled by a marketing research company.
The population characteristics of these two-source datasets are naturally different.
The MAR assumption does not hold in such cases, and the methods that do not consider nonnegligible missingness may yield severely biased estimates.
In the field of labor economics, a probit selection model is used to adjust the selection bias that occurs in a single dataset with missing variables (Heckman, 1974, 1979).
Various methods have been proposed in addition to Heckman's probit selection model (e.g., Qin et al., 2002); however, they are not directly applicable to the data fusion or data combination problems.

In latent variable modeling, the exploratory factor analysis model (EFAM; Cudeck, 2000; Kamakura and Wedel, 2000) and latent class model (Kamakura and Wedel, 2003) have been applied to the data combination problem. These models assume latent factors and latent classes behind observations.
However, as they assume an MAR  situation,  results may be affected by selection bias in a non-MAR (NMAR) situation.Moreover, these models assume a linear mapping from latent variables to observed variables and require parametric estimation. Deviations from model assumptions result in poor data fitting.

To prevent such parametric model assumptions, we extend the latent variable model to a nonparametric model using the Gaussian process (GP) applied in the field of machine learning.
The GP is a probability distribution that generates a random function, and the function values obtained from the GP can be used to capture the nonlinear relationships between variables.
The GP latent variable model (GPLVM) has been proposed as a dimensionality reduction method for observed data (Lawrence, 2004, 2005). This method applies the GP to principal component analysis and consists of unsupervised learning.
Furthermore, a Bayesian approach has been proposed to estimate the parameters of this method (Titsias and Lawrence, 2010).
In this approach, the similarities between a unit's latent variables are used to construct the distribution of observed variables because the Gram matrix with latent variables is used as the covariance function.
This method uses the distance between latent variables. It is analogous to incorporating the relational data of latent variables into the model  and similar to techniques such as matching between latent variables and graph embedding.

GP modeling has been applied to general methods such as regression and classification models, and promising nonparametric Bayesian approaches  have been developed (Williams and Rasmussen, 1996; Williams and Barber, 1998).
Moreover, the log Gaussian Cox process (M{\o}ller et al., 1998) and discriminative GPLVM (Urtasun and Darrell, 2007) have been used for discrete and continuous variables.
Propensity score matching with the GP has been proposed to solve the data combination problem. This method estimates the matching weights that capture the nonlinear relationship between covariates (Vegetabile et al., in press).
As another method of covariate dimension reduction, the combination of the GPLVM with a Weibull proportional hazards model has been proposed to extract a low-dimensional structure of covariates from survival data (Barrett and Coolen, 2016).

In  this study, we propose a data fusion method that does not assume that datasets to be fused are homogenous. The method uses a GPLVM for NMAR missing data and assumes that the variables of concern and the probability of being missing depend on latent variables.
Furthermore, we propose a GP data combination model  (GPDCM) for NMAR missing data. This model can predict the values of missing outcome variables because the latent variables of all units are estimated even if outcome variables are missing.
Our method assumes that the probability of being missing depends on latent variables; hence, it can adjust the selection bias in the data fusion problem.
Moreover, the GPDCM is a probability generation model in a unified framework with the GP, which estimates the latent variables that are assumed behind observed variables and generates discrete and continuous observed values from a latent function of the GP.
We show the usefulness of the GPDCM using a simulation study and real-world data analyses and comparing it with several other methods.

Table \ref{dcm} lists the different data combination methods.
% 表の挿入
\begin{table}[tbhp]
 \caption{Data combination methods based on latent variable modeling}% {}内に表題を書く
 \label{dcm}
 \begin{center}
  \begin{tabular}{lll}
    \noalign{\smallskip}\hline\noalign{\smallskip}
     Model  &  Missing-data mechanism  &    \\
    \noalign{\smallskip}\cline{2-3}\noalign{\smallskip}
       &  MAR  &  NMAR  \\
    \noalign{\smallskip}\hline\noalign{\smallskip}
     non-GP  &  EFAM to deal with missing data   &  LVM-NMAR  \\
             &  (Kamakura and Wedel, 2000)  &  (proposed method) \vspace{2mm} \\
     GP      &  GPDCM-MAR  &  GPDCM-NMAR  \\
             &  (proposed method)  &  (proposed method)  \\
    \noalign{\smallskip}\hline\noalign{\smallskip}
  \end{tabular}
 \end{center}
\end{table}

\section{Proposed data combination method}
We propose the GPDCM for mixed discrete and continuous data with NMAR missingness to integrate multiple-source datasets into a single-source dataset.
In this section, we first describe the missing-data mechanism, which is crucial but frequently neglected in the data combination problem.
Next, we describe mixed data in the exponential family. Then, the GPDCM for NMAR missing data is formulated, and the estimation algorithm is presented.

\subsection{Data combination problem}
Let $\boldsymbol{Y}=(Y_1, Y_2, \cdots, Y_p)^{'}$ and $\boldsymbol{X}$ be $p$-dimensional random vectors. Additionally, suppose that the distribution of $\boldsymbol{Y}$ permits a probability density function, $p(\boldsymbol{y}|\boldsymbol{\vartheta})$, where $\boldsymbol{y}=(y_1, y_2, \cdots, y_p)^{'}\in  \mathbb{R}^p$ and $\boldsymbol{\vartheta}\in \mathbb{R}^q$ denote an observable value vector and a parameter of the distribution, respectively.
Let $\boldsymbol{M}=(M_1, M_2, \cdots, M_p)^{'}$ be a vector of missing indicators, as follows:
\begin{align}
\begin{split}
M_j&=\left\{
  \begin{array}{cl}
    1   &  (Y_j \mathrm{\ is \ observed})  \\
    0   &  (Y_j \mathrm{\ is \ missing})  \\
  \end{array}
\right. \\
(j&=1, 2, \cdots, p).
\end{split}
\end{align}
Let $\boldsymbol{m}^{(\ell)}=(m_1, m_2, \cdots, m_p)^{'}$ $(\ell=1, 2, \cdots, L)$ be any value that $\boldsymbol{M}$ can take, where $L\le 2^p$ is the number of missing patterns in $\boldsymbol{Y}$.
%When all elements of $\boldsymbol{Y}$ are observed, the missing pattern is denoted by $\boldsymbol{m}^{(1)}=(1, 1, \cdots, 1)$.
According to the missing pattern, $\boldsymbol{M}=\boldsymbol{m}^{(\ell)}$, $\boldsymbol{Y}^{(\ell)}$ denotes a set of observed values and $\boldsymbol{Y}^{(-\ell)}$ denotes a set of missing values.
Then, the missing data are  represented by $\boldsymbol{Y}=(\boldsymbol{Y}^{(\ell)'}, \boldsymbol{Y}^{(-\ell)'})^{'}$ and $\boldsymbol{y}=(\boldsymbol{y}^{(\ell)'}, \boldsymbol{y}^{(-\ell)'})^{'}$.
%In Figure \ref{datadef}, the missing pattern of outcome variables is defined as follows:
%\begin{align}
%\begin{split}
%\boldsymbol{m}^{(1)}&=(1,\cdots,1,0,\cdots 0)=\boldsymbol{m} \\
%\boldsymbol{m}^{(2)}&=(0,\cdots,0,1,\cdots 1)=1-\boldsymbol{m}.
%\end{split}
%\end{align}

To focus on parameter $\boldsymbol{\vartheta}$, the selection model decomposes the joint distribution of $(\boldsymbol{X}, \boldsymbol{Y}, \boldsymbol{M})$ as follows:
\begin{align}
P(\boldsymbol{X}, \boldsymbol{Y}, \boldsymbol{M}|\boldsymbol{\vartheta},\boldsymbol{\varphi})=p(\boldsymbol{x},\boldsymbol{y}|\boldsymbol{\vartheta})P(\boldsymbol{M}=\boldsymbol{m}^{(\ell)}|\boldsymbol{X}=\boldsymbol{x},\boldsymbol{Y}=\boldsymbol{y};\boldsymbol{\varphi}),
\label{smodel}
\end{align}
where $\boldsymbol{\varphi}\in  \mathbb{R}^q$ is a nuisance parameter.

Based on the functional form of the conditional probability, $P(\boldsymbol{M}=\boldsymbol{m}^{(\ell)}|\boldsymbol{X}=\boldsymbol{x},\boldsymbol{Y}=\boldsymbol{y};\boldsymbol{\varphi})$, the missing-data mechanism is classified into three type of missing data, MCAR (missing completely at random), MAR (missing at random), and NMAR (not missing at random) (Rubin, 1976).

Selection model (\ref{smodel}) use used to obtain the joint distribution of observable data $\boldsymbol{X}, \boldsymbol{Y}^{(\ell)}$ and missing pattern $\boldsymbol{M}$ as follows:
\begin{align}
\begin{split}
&P(\boldsymbol{X}=\boldsymbol{x},\boldsymbol{Y}^{(\ell)}=\boldsymbol{y}^{(\ell)}, \boldsymbol{M}=\boldsymbol{m}^{(\ell)}|\boldsymbol{\vartheta},\boldsymbol{\varphi}) \\
&=\int_{\boldsymbol{y}^{(-\ell)}} P(\boldsymbol{X}=\boldsymbol{x}, \boldsymbol{Y}=\boldsymbol{y}, \boldsymbol{M}=\boldsymbol{m}^{(\ell)}|\boldsymbol{\vartheta},\boldsymbol{\varphi})d\boldsymbol{y}^{(-\ell)} \\
&=\int_{\boldsymbol{y}^{(-\ell)}} p(\boldsymbol{x},\boldsymbol{y}|\boldsymbol{\vartheta})P(\boldsymbol{M}=\boldsymbol{m}^{(\ell)}|\boldsymbol{X}=\boldsymbol{x},\boldsymbol{Y}=\boldsymbol{y};\boldsymbol{\varphi})d\boldsymbol{y}^{(-\ell)}.
\label{miss}
\end{split}
\end{align}
This model can also represent a discrete distribution by replacing the integral with the summation.
In the case of MAR, joint distribution (\ref{miss}) is as follows:
\begin{align}
\begin{split}
p(\boldsymbol{x},\boldsymbol{y}^{(\ell)}|\boldsymbol{\vartheta})P(\boldsymbol{M}=\boldsymbol{m}^{(\ell)}|\boldsymbol{X}=\boldsymbol{x},\boldsymbol{Y}^{(\ell)}=\boldsymbol{y}^{(\ell)};\boldsymbol{\varphi}).
\end{split}
\end{align}
Here, the likelihood inference obtained using only the observed data with $p(\boldsymbol{x},\boldsymbol{y}^{(\ell)}|\boldsymbol{\vartheta})$ creates the estimator.
Although we can neglect the missing indicators, $\boldsymbol{M}$, in MAR missingness, it is necessary to assume a probability model for $\boldsymbol{M}$ in NMAR missingness.

We use a shared-parameter model framework to deal with NMAR missingness in the data combination problem (e.g., Follman and Wu, 1995); this model introduces latent variables.
Let $\boldsymbol{Z}=(Z_1, Z_2, \cdots, Z_d)^{'}$ be a covariate variable. We assume that the conditional independence assumption of $\boldsymbol{Y}$ and $\boldsymbol{M}$ is as follows:
\begin{align}
(\boldsymbol{X}, \boldsymbol{Y})\Perp \boldsymbol{M} \ | \ \boldsymbol{Z}.
\end{align}
Under this condition, the shared-parameter model decomposes the joint distribution of $(\boldsymbol{X}, \boldsymbol{Y}, \boldsymbol{M})$ using latent variables as follows:
\begin{align}
P(\boldsymbol{X}, \boldsymbol{Y}, \boldsymbol{M})=\int_{\boldsymbol{z}}p(\boldsymbol{x}, \boldsymbol{y}|\boldsymbol{z})P(\boldsymbol{M}=\boldsymbol{m}^{(\ell)}|\boldsymbol{z})p(\boldsymbol{z})d\boldsymbol{z},
\label{spm_f}
\end{align}
The joint distribution of observed data is as follows:
\begin{align}
\begin{split}
&P(\boldsymbol{X}, \boldsymbol{Y}^{(\ell)}, \boldsymbol{M}) \\
&=\int_{\boldsymbol{y}^{(-\ell)}}\int_{\boldsymbol{z}}p(\boldsymbol{x}, \boldsymbol{y}^{(\ell)}|\boldsymbol{z})p(\boldsymbol{y}^{(-\ell)}|\boldsymbol{z})P(\boldsymbol{M}=\boldsymbol{m}^{(\ell)}|\boldsymbol{z})p(\boldsymbol{z})d\boldsymbol{z}d\boldsymbol{y}^{(-\ell)} \\
&=\int_{\boldsymbol{z}}p(\boldsymbol{x}, \boldsymbol{y}^{(\ell)}|\boldsymbol{z})P(\boldsymbol{M}=\boldsymbol{m}^{(\ell)}|\boldsymbol{z})p(\boldsymbol{z})\left\{\int_{\boldsymbol{y}^{(-\ell)}}p(\boldsymbol{y}^{(-\ell)}|\boldsymbol{z})\boldsymbol{y}^{(-\ell)}\right\}d\boldsymbol{z} \\
&=\int_{\boldsymbol{z}}p(\boldsymbol{x}, \boldsymbol{y}^{(\ell)}|\boldsymbol{z})P(\boldsymbol{M}=\boldsymbol{m}^{(\ell)}|\boldsymbol{z})p(\boldsymbol{z})d\boldsymbol{z},
\label{spm}
\end{split}
\end{align}
and in the case of MAR,
\begin{align}
P(\boldsymbol{X}, \boldsymbol{Y}^{(\ell)}, \boldsymbol{M})=\int_{\boldsymbol{z}}p(\boldsymbol{x}, \boldsymbol{y}^{(\ell)}|\boldsymbol{z})p(\boldsymbol{z})d\boldsymbol{z}.
\end{align}
These formulations are the basis of data combination models, and they can be used according to the missing pattern.
As shown in Figure \ref{datadef}, the data combination approach consists of two missing patterns, $L=2$, where $(\ell, \ell^{'})=(1, 2)$ and $(2, 1)$ for outcome variables $Y_1$ and $Y_2$, respectively.
Here, two missing indicators are denoted by $\boldsymbol{m}^{(\ell)}$ and $\boldsymbol{m}^{(\ell^{'})}=\boldsymbol{1}_p-\boldsymbol{m}^{(\ell)}$.

\begin{figure}[tbhp]
  \begin{center}
    \includegraphics[scale=0.38]{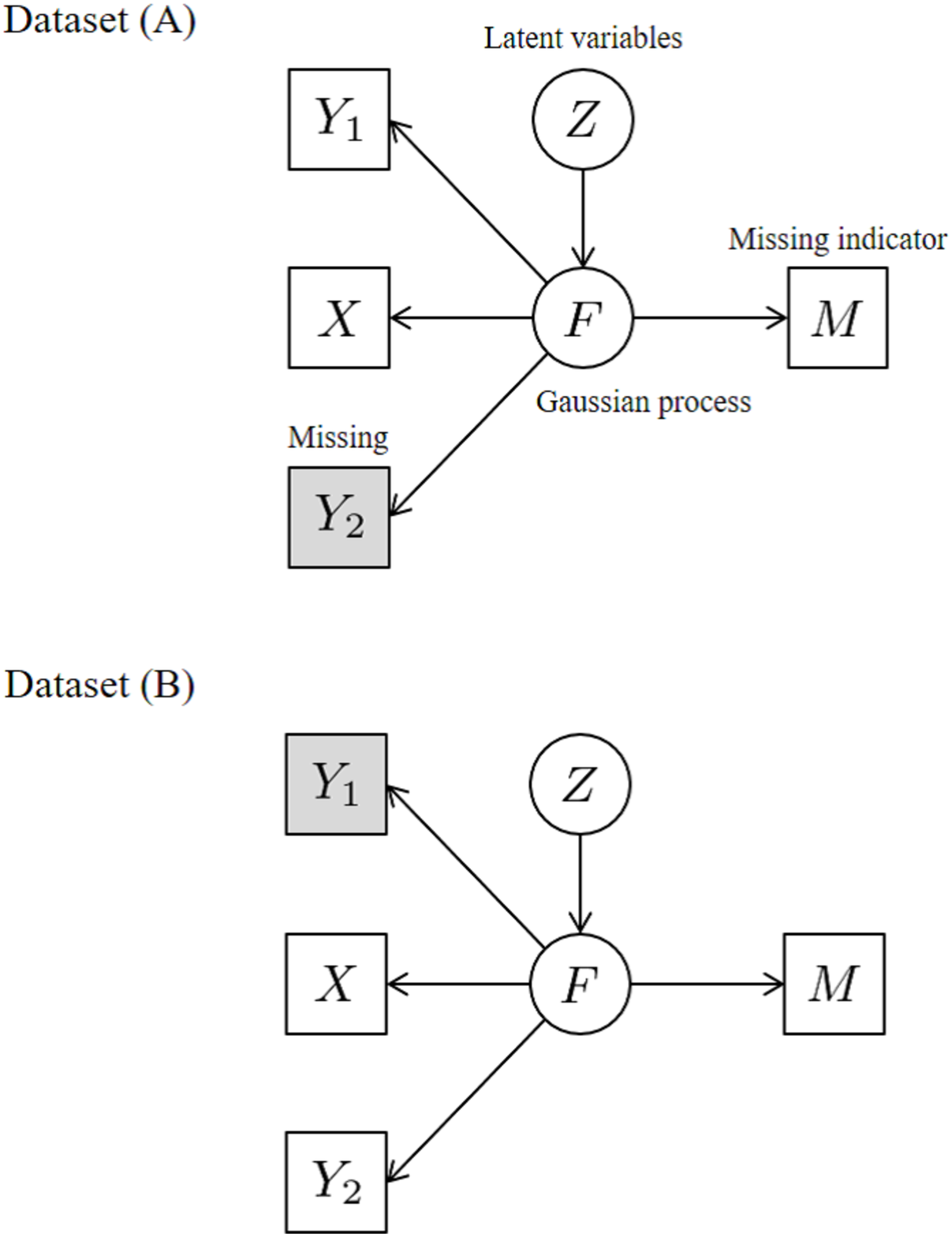}
  \end{center}
  \caption{Overview of data combination model for NMAR missing data}%・ｽ^・ｽC・ｽg・ｽ・ｽ・ｽ・ｽ・ｽL・ｽ・ｽ
  \label{nmar_model}
\end{figure}

The data combination approach considers the following conditional independence assumption:
\begin{align}
\begin{split}
\boldsymbol{Y}_1&\Perp \boldsymbol{Y}_2 \ | \ \boldsymbol{Z}, \\
P(\boldsymbol{Y}_1,\boldsymbol{Y}_2|\boldsymbol{Z})&=P(\boldsymbol{Y}_1|\boldsymbol{Z})P(\boldsymbol{Y}_2|\boldsymbol{Z}).
\label{ci}
\end{split}
\end{align}
When this assumption holds, each unit of the same covariate does not necessarily have the same relationship between outcome variables.
We use conditional independence assumption (\ref{ci}) to express the joint distribution of observable data, $\boldsymbol{X}, \boldsymbol{Y}^{(\ell)}_1, \boldsymbol{Y}^{(\ell^{'})}_2$, and missing pattern $\boldsymbol{M}$ as follows:\begin{align}
\begin{split}
P&(\boldsymbol{X}, \boldsymbol{Y}^{(\ell)}_1, \boldsymbol{Y}^{(\ell^{'})}_2, \boldsymbol{M}) \\
=&\int_{\boldsymbol{z}}p(\boldsymbol{x},\boldsymbol{y}^{(\ell)}_1,\boldsymbol{y}^{(\ell^{'})}_2|\boldsymbol{z})P(\boldsymbol{M}=\boldsymbol{m}^{(\ell)}|\boldsymbol{z})p(\boldsymbol{z})d\boldsymbol{z} \\
=&\int_{\boldsymbol{z}}p(\boldsymbol{x}|\boldsymbol{z})p(\boldsymbol{y}^{(\ell)}_1|\boldsymbol{z})p(\boldsymbol{y}^{(\ell^{'})}_2|\boldsymbol{z})P(\boldsymbol{M}=\boldsymbol{m}^{(\ell)}|\boldsymbol{z})p(\boldsymbol{z})d\boldsymbol{z} \\
=&\int_{\boldsymbol{z}}\prod_{i:m_i=1}p(x_i|\boldsymbol{z}_i)p(y^{(\ell)}_{1i}|\boldsymbol{z}_i)p(m_i=1|\boldsymbol{z}_i) \\
&\prod_{i:m_i=0}p(x_i|\boldsymbol{z}_i)p(y^{(\ell^{'})}_{2i}|\boldsymbol{z}_i)p(m_i=0|\boldsymbol{z}_i)p(\boldsymbol{z})d\boldsymbol{z}.
\label{dc_nmar}
\end{split}
\end{align}
The likelihood is represented by shared-parameter model (\ref{spm}) because latent variable $\boldsymbol{Z}$ is unobserved.
In the case of MAR, joint distribution (\ref{dc_nmar}) is expressed as follows (Kamakura and Wedel, 2000):
\begin{align}
\begin{split}
&P(\boldsymbol{X}, \boldsymbol{Y}^{(\ell)}_1, \boldsymbol{Y}^{(\ell^{'})}_2, \boldsymbol{M}) \\
&=\int_{\boldsymbol{z}}\prod_{i:m_i=1}p(x_i|\boldsymbol{z}_i)p(y^{(\ell)}_{1i}|\boldsymbol{z}_i)\prod_{i:m_i=0}p(x_i|\boldsymbol{z}_i)p(y^{(\ell^{'})}_{2i}|\boldsymbol{z}_i)p(\boldsymbol{z})d\boldsymbol{z}.
\label{dc_mar}
\end{split}
\end{align}
Both likelihoods with the latent variables for MAR and NMAR missing data can be used to solve the data combination problem.
We replace the latent variables in the data combination model with GP latent functions, as shown in Figure \ref{nmar_model}.
These functions are used to generate observed data and missing indicators for two-source datasets (A) and (B) in Figure \ref{datadef}.

\subsection{GPLVM for mixed discrete and continuous data}
The observed values, $y_{ij}$ or $x_{ij}$, are random values that have the following probability density function in the exponential family:
\begin{align}
\displaystyle p_j(y_{ij}|\theta_{ij}, \phi_j)=\exp\left\{\frac{y_{ij}\theta_{ij}-\alpha_j(\theta_{ij})}{\phi_j}+\beta_j(y_{ij}, \phi_j)\right\},
\end{align}
where $\theta_{ij}$ and $\phi_j$ are the canonical and dispersion parameters, respectively, and $\alpha_j(\cdot)$ and $\beta_j(\cdot)$ are functions that depend on the particular distribution function of the $j$-th variable.
The exponential family can represent discrete and continuous distributions. Table \ref{expo} shows a few representative distributions.
Observed variables are determined by changing the canonical link function of each distribution. This function relates a canonical parameter to the expectation of a random variable. The canonical links for the normal, Bernoulli, and Poisson distributions are the identity link, logit link, and log link function, respectively.
\begin{table}[tbhp]
 \caption{Representative discrete and continuous distributions in the univariate exponential family}% {}内に表題を書く
 \label{expo}
 \begin{center}
  \begin{tabular}{lccl}
    \noalign{\smallskip}\hline\noalign{\smallskip}
       Distribution  &  $p(y)$  &  Domain  &  Link function  \\
    \noalign{\smallskip}\hline\noalign{\smallskip}
        Bernoulli: $\mathcal{B}(\pi)$   &  $\pi^y(1-\pi)^{1-y}$  &  $[0, 1]$  &  $\displaystyle\theta=\ln\left(\frac{\pi}{1-\pi}\right)$  \\
        Poisson: $\mathcal{P}(\mu)$   &  $\displaystyle\frac{e^{-\mu}\mu^y}{y!}$  &  $(0, \infty)$  &  $\theta=\ln(\mu)$  \\
        Normal: $\mathcal{N}(\mu, \sigma^2)$   &  $\displaystyle\frac{1}{\sqrt{2\pi\sigma}}\exp\left\{\frac{-(y-\mu)^2}{2\sigma^2}\right\}$  &  $(-\infty, \infty)$  &  $\theta=\mu$  \\
    \noalign{\smallskip}\hline\noalign{\smallskip}
  \end{tabular}
 \end{center}
\end{table}

In the latent variable model, canonical parameter vector $\boldsymbol{\theta}_i$ contains low-dimensional latent variables, $\boldsymbol{z}_i$, and it is defined as follows (Kamakura and Wedel, 2000; Wedel and Kamakura, 2001):
\begin{align}
\boldsymbol{\theta}_i=\boldsymbol{\lambda}_0+\boldsymbol{\varLambda}\boldsymbol{z}_i,
\label{cano}
\end{align}
where $\boldsymbol{\lambda}_0$ and $\boldsymbol{\varLambda}$ denote an intercept vector and the factor loading matrix of latent variables, respectively.
%Here, an $n \times 1$ vector of ones and the transposed matrix of a matrix $\boldsymbol{X}$ are denoted by $\boldsymbol{1}_n$ and $\boldsymbol{X}^{'}$, respectively.
This latent variable model represents a linear relationship between latent variables and observed variables, and it is used for the distributions listed in Table \ref{expo}.

When $\boldsymbol{z}_i$ is mapped to a high-dimensional feature space, observed vector $\boldsymbol{y}$ is defined by design matrix $\boldsymbol{\varPhi}_{\boldsymbol{z}}$, which consists of future vectors $\phi_{\boldsymbol{z}}(\boldsymbol{z}_i)$, as follows:
\begin{align}
\boldsymbol{y}=\boldsymbol{\varPhi}_{\boldsymbol{z}}\boldsymbol{w},
\end{align}
where $\boldsymbol{w}$ is a high-dimensional loading vector.
Assuming that $\boldsymbol{w}\sim \mathcal{N}(\boldsymbol{0}, \sigma^2\boldsymbol{I}_D)$, $\boldsymbol{y}$ is expressed as follows:
\begin{align}
\boldsymbol{y}\sim \mathcal{N}(\boldsymbol{0}, \sigma^2\boldsymbol{\varPhi}_{\boldsymbol{z}}\boldsymbol{\varPhi}^{'}_{\boldsymbol{z}}).
\end{align}
The covariance of this normal distribution is the Gram matrix, $\boldsymbol{K}_{\boldsymbol{z}, \boldsymbol{\tau}}=\boldsymbol{\varPhi}_{\boldsymbol{z}}\boldsymbol{\varPhi}^{'}_{\boldsymbol{z}}$.

By estimating the kernel with the dimension of the number of units in the data, we can generate $boldsymbol{y}$ without estimating the high-dimensional loading vector $\boldsymbol{w}$, and avoid the curse of dimensionality.

Let $f_{(j)}(z_i)$ $(i=1, 2, \cdots, n; j=1, 2 , \cdots, p)$ be a latent function value of $d$-dimensional latent variables.
Then, $f_{(j)}(z_i)$ is obtained from the GP and denoted by $\boldsymbol{f}_{(j)}\sim \mathcal{GP}(\boldsymbol{\mu}, \boldsymbol{K}_{\boldsymbol{z}, \boldsymbol{\tau}})$, where $\boldsymbol{\mu}=(\mu(\boldsymbol{z}_1), \mu(\boldsymbol{z}_2), \cdots, \mu(\boldsymbol{z}_n))$ is a mean vector and $\boldsymbol{K}_{\boldsymbol{z}, \boldsymbol{\tau}}=(k_{\boldsymbol{z}, \boldsymbol{\tau}}(\boldsymbol{z}_i,\boldsymbol{z}^{'}_j))$ is a kernel covariate matrix obtained using hyperparameter vector $\boldsymbol{\tau}$.
When the distribution of the latent variables is $\boldsymbol{z}_i\sim \mathcal{N}(\boldsymbol{0}_d, \boldsymbol{I}_d)$, a latent function vector is generated by $\boldsymbol{f}_{(j)}\sim \mathcal{N}(\boldsymbol{0}, \boldsymbol{K}_{\boldsymbol{z}, \boldsymbol{\tau}})$, and latent function matrix $\boldsymbol{F}$ is defined as follows:
\begin{align}
\begin{split}
\boldsymbol{f}_{(j)}&=(f_{1j}, f_{2j}, \cdots, f_{nj})^{'}=(f_{(j)}(\boldsymbol{z}_1), f_{(j)}(\boldsymbol{z}_2), \cdots, f_{(j)}(\boldsymbol{z}_n))^{'}, \\
\boldsymbol{f}_{i}&=(f_{i1}, f_{i2}, \cdots, f_{ip})^{'}=(f_{(1)}(\boldsymbol{z}_i), f_{(2)}(\boldsymbol{z}_i), \cdots, f_{(p)}(\boldsymbol{z}_i))^{'}.
\end{split}
\end{align}
This GP latent function vector, $\boldsymbol{f}_i$, is treated as the weighted latent variables, $\boldsymbol{\varLambda}\boldsymbol{z}_i$, estimated from observed variables.
$\boldsymbol{f}_i$ is used to reformulate canonical parameter (\ref{cano}) as follows:
\begin{align}
\boldsymbol{\theta}_i=\boldsymbol{\lambda}_0+\boldsymbol{f}_i.
\label{canof}
\end{align}
Unlike latent variable model (\ref{cano}), the model with the GP assumes that there is a nonlinear relationship between the latent variables and observed variables. This model is used for the distributions listed in Table \ref{expo}.
Suppose that the observed values, $x_{ij}$, $y_{ij1}$, and $y_{ij2}$, follow one of the following three distributions that are modeled by the GP:
\begin{itemize}

\item Bernoulli: $\mathcal{B}(\pi)$, where $\displaystyle\pi=\frac{1}{1+\exp\left\{-\left(\lambda_{0ij}+f_{(j)}(\boldsymbol{z}_i)\right)\right\}}$.

\item Poisson: $\mathcal{P}(\mu)$, where $\mu=\exp\left(\lambda_{0ij}+f_{(j)}(\boldsymbol{z}_i)\right)$.

\item Normal: $\mathcal{N}(\mu, \sigma^2)$, where $\mu=\lambda_{0ij}+f_{(j)}(\boldsymbol{z}_i)$.

\end{itemize}
These distributions are used in various methods such as GP regression, GP Poisson regression, and GP discriminant models.
Canonical parameter (\ref{canof}) derived from GP latent functions can be used for several types of distributions with discrete and continuous variables.

\subsection{GPDCM for NMAR missing data}
The sets of observed data and parameters are $\mathcal{D}=\bigl\{(\boldsymbol{x}_i, \boldsymbol{y}_{i1}, \boldsymbol{y}_{i2})^{n}_{i=1}\bigl\}$ and $\mathcal{H}=\{\boldsymbol{f}, \boldsymbol{z}, \boldsymbol{\tau}\}$.
First, we consider the case of MAR missing data, which do not depend on missing variables.
According to Figure \ref{datadef} and joint distribution (\ref{dc_mar}), when the observed data are given, the likelihood of the GPDCM in MAR missingness is as follows:
\begin{align}
\begin{split}
p(\mathcal{D})=&\iint p(\boldsymbol{x},\boldsymbol{y}_1,\boldsymbol{y}_2|\boldsymbol{f})p(\boldsymbol{f}|\boldsymbol{z},\boldsymbol{\tau})p(\boldsymbol{z})d\boldsymbol{z}d\boldsymbol{f} \\
=&\iint \prod^n_{i=1}\Bigl\{p(\boldsymbol{x}_{i}|\boldsymbol{f}_{xi},\phi_{x})p(\boldsymbol{y}_{i1}|\boldsymbol{f}_{y_1i},\phi_{y_1})\Bigl\}^{m_i} \\
&\Bigl\{p(\boldsymbol{x}_{i}|\boldsymbol{f}_{xi},\phi_{x})p(\boldsymbol{y}_{i2}|\boldsymbol{f}_{y_2i},\phi_{y_2})\Bigl\}^{1-m_i} \\
&\prod^p_{j=1}p(\boldsymbol{f}_{(j)}|\boldsymbol{K}_{\boldsymbol{z},\boldsymbol{\tau}_{(j)}})\prod^n_{i=1}p(\boldsymbol{z}_i)d\boldsymbol{z}d\boldsymbol{f},
\end{split}
\end{align}
%where $\boldsymbol{f}_i=(\boldsymbol{f}^{'}_{xi},\boldsymbol{f}^{'}_{y_1i},\boldsymbol{f}^{'}_{y_2i})^{'}$.
The proposed GPDCM-MAR can be applied to mixed data to capture the nonlinear relationship between variables.

Next, we consider the case of NMAR missing data, which depend on missing variables.
It is necessary to assume a structure for missing indicator $m_i$ to express joint distribution (\ref{dc_nmar}).
We assume the following probit model to determine $m_i$ depending on latent variables:
\begin{align}
\begin{split}
u_i&=f_{(p+1)}(\boldsymbol{z}_i)+\varepsilon_i, \ \varepsilon_i\sim \mathcal{N}(0, 1), \\
m_i&=\left\{
  \begin{array}{cc}
    1   &  (u_i > 0)  \\
    0   &  (u_i \le  0)  \\
  \end{array}
\right., \\
(i&=1, 2, \cdots, n).
\label{mi1}
\end{split}
\end{align}
In this model, the probabilities of being missing are
\begin{align}
\begin{split}
p(m_i=1|\boldsymbol{z}_i)&=\boldsymbol{\varPhi}(f_{(p+1)}(\boldsymbol{z}_i)), \\
p(m_i=0|\boldsymbol{z}_i)&=1-\boldsymbol{\varPhi}(f_{(p+1)}(\boldsymbol{z}_i)).
\label{mi2}
\end{split}
\end{align}
where $\varPhi(\cdot)$ is the cumulative distribution function of the standard normal distribution.
In the probit model without the GP, latent variable $f_{(p+1)}(\boldsymbol{z}_i)$ is replaced by $\boldsymbol{z}^{'}_i\boldsymbol{\lambda}_{(p+1)}$.

Based on the above, when the observed data are given, the likelihood of the GPDCM in NMAR missingness is expressed as follows:
\begin{align}
\begin{split}
p(\mathcal{D},\boldsymbol{m})=&\iint p(\boldsymbol{x},\boldsymbol{y}_1,\boldsymbol{y}_2,\boldsymbol{m}|\boldsymbol{f})p(\boldsymbol{f}|\boldsymbol{z},\boldsymbol{\tau})p(\boldsymbol{z})d\boldsymbol{z}d\boldsymbol{f} \\
=&\iint\prod^n_{i=1}\Bigl\{p(\boldsymbol{x}_{i}|\boldsymbol{f}_{xi},\phi_{x})p(\boldsymbol{y}_{i1}|\boldsymbol{f}_{y_1i},\phi_{y_1})p(m_i=1|\boldsymbol{f}_{mi})\Bigl\}^{m_i} \\
&\Bigl\{p(\boldsymbol{x}_{i}|\boldsymbol{f}_{xi},\phi_{x})p(\boldsymbol{y}_{i2}|\boldsymbol{f}_{y_2i},\phi_{y_2})p(m_i=0|\boldsymbol{f}_{mi})\Bigl\}^{1-m_i} \\
&\prod^p_{j=1}p(\boldsymbol{f}_{(j)}|\boldsymbol{K}_{\boldsymbol{z},\boldsymbol{\tau}_{(j)}})\prod^n_{i=1}p(\boldsymbol{z}_i)d\boldsymbol{z}d\boldsymbol{f}.
\end{split}
\end{align}
As in the case of the proposed GPDCM-MAR, the proposed GPDCM-NMAR can be applied to mixed observed data to capture the nonlinear relationship between variables using the latent GP function.

\section{Markov chain Monte Carlo algorithm}
The missing observations and parameters are alternatively sampled using the random walk Metropolis--Hastings (RWMH) algorithm, which is a Markov chain Monte Carlo (MCMC) algorithm.
The algorithm of GPDCM-NMAR is as follows:
\begin{description}
  \item[Step1:]Draw latent variables (update $\boldsymbol{z}_i$).
  \item[Step2:]Draw kernel parameters (update $\boldsymbol{\tau}_{(j)}$).
  \item[Step3:]Draw GP latent function (update $\boldsymbol{f}_{(j)}$).
  \item[Step4:]Draw intercept of outcome variables including missing observations (update $\lambda_{0p^{\mathrm{out}}_j}$).
  \item[Step5:]Draw hyperparameters of observations (update $\sigma$).
  \item[Step6:]Draw missing observations (update $\boldsymbol{y}_1,\boldsymbol{y}_2$).
\end{description}

The proposed method estimates the GP latent functions and each parameter and uses them to predict the missing values of outcome variables.
When estimated parameter set $\hat{\mathcal{H}}$ is given, the posterior predictive distribution of missing observations is as follows:
\begin{align}
\begin{split}
p(\boldsymbol{y}_1,\boldsymbol{y}_2|\mathcal{D},\boldsymbol{m})=&\iint p(\boldsymbol{y}_1,\boldsymbol{y}_2|\boldsymbol{f},\hat{\mathcal{H}},\mathcal{D},\boldsymbol{m})p(\boldsymbol{f}|\boldsymbol{z},\hat{\mathcal{H}},\mathcal{D},\boldsymbol{m}) \\
&p(\boldsymbol{z}|\hat{\mathcal{H}},\mathcal{D},\boldsymbol{m})p(\hat{\mathcal{H}}|\mathcal{D},\boldsymbol{m})d\boldsymbol{z}d\boldsymbol{f}.
\end{split}
\end{align}
Once the GP latent functions are estimated for each unit, the missing observations are obtained from the binomial, Poisson, or normal distributions. The distribution to be used depends on the type of outcome variables.

The kernel function, $k$, of the covariate matrix is expressed as the following Gaussian kernel, which is common in machine learning:
\begin{align}
k(\boldsymbol{z},\boldsymbol{z}^{'})=\tau_1 \exp \left(-\frac{||\boldsymbol{z}-\boldsymbol{z}^{'}||^2}{2\tau_2}\right), \ \boldsymbol{\tau}=(\tau_1, \tau_2)^{'}.
\end{align}
$\tau_1$ and $\tau_2$ are the hyperparameters of the kernel function.
The Gram matrix obtained from this kernel function is used to generate the GP latent function vector from $\boldsymbol{f}_{(j)}\sim \mathcal{N}(\boldsymbol{0}_n, \boldsymbol{K}_{\boldsymbol{z},\boldsymbol{\tau}_{(j)}})$.
For computational efficiency, this latent function value is defined as follows:
\begin{align}
\boldsymbol{f}_{(j)}=\boldsymbol{L}\boldsymbol{\eta}_{(j)},
\end{align}
where $\boldsymbol{L}$ is a lower triangular matrix with real and positive diagonal entries obtained from the Cholesky decomposition, $\boldsymbol{K}_{\boldsymbol{z},\boldsymbol{\tau}_{(j)}}=\boldsymbol{L}\boldsymbol{L}^{'}$.
In addition, $\boldsymbol{\eta}_{(j)}$ is an isotropic unit normal variate generated from $\boldsymbol{\eta}_{(j)}\sim \mathcal{N}(\boldsymbol{0}_n, \boldsymbol{I}_n)$, where $\boldsymbol{I}_n$ is the identity matrix.
$\boldsymbol{\eta}_{(j)}$ is independent of the hyperparameters.

The hyperparameters for the fixed latent GP function are typically updated using the following conditional posterior distribution:
\begin{align}
p(\boldsymbol{\tau}_{(j)}|\boldsymbol{f}_{(j)})\propto \mathcal{N}(\boldsymbol{0}, \boldsymbol{K}_{\boldsymbol{z},\boldsymbol{\tau}_{(j)}})p(\boldsymbol{\tau}_{(j)}).
\end{align}

However, the above parameter update program limits the number of Markov chains for which hyperparameters can be updated for a fixed latent GP function, making the search for the posterior distribution time-consuming (Murray and Adams, 2010).
Therefore, instead of the above parameter update algorithm, in this study we follow the following RWMH algorithm to update both $\boldsymbol{\tau}_{(j)}$ and $\boldsymbol{f}_{(j)}$.

\begin{algorithm}
\renewcommand{\thealgorithm}{}
\caption{{\bf RWMH algorithm}}
\label{}
\begin{algorithmic}[1]
\renewcommand{\algorithmicrequire}{\textbf{Input:}}
\renewcommand{\algorithmicensure}{\textbf{Output:}}
\REQUIRE $\tau_{(j)}$, $\boldsymbol{f}_{(j)}$ and other parameters in the last iteration.
\ENSURE  Next $\tau_{(j)}$, $\boldsymbol{f}_{(j)}$
\STATE Cholesky decomposition: $\boldsymbol{K}_{\boldsymbol{z}}=\boldsymbol{L}\boldsymbol{L}^{'}$
\STATE Solve for $\mathcal{N}(\boldsymbol{0}_n, \boldsymbol{I}_n)$ variate: $\boldsymbol{\eta}_{(j)}=\boldsymbol{L}^{-1}\boldsymbol{f}_{(j)}$
\STATE Propose $\tau^{'}_{(j)}\sim p(\tau^{'}_{(j)}; \tau_{(j)})$
\STATE Compute implied values: $\boldsymbol{f}^{'}_{(j)}=\boldsymbol{L}\boldsymbol{\eta}_{(j)}$
\STATE Draw $u\sim \mathcal{U}(0, 1)$
\IF{$\displaystyle u < u_{\tau_{(j)}}=\frac{p(\boldsymbol{y}|\boldsymbol{f}^{'}_{(j)}, \tau^{'}_{(j)})p(\tau^{'}_{(j)})}{p(\boldsymbol{y}|\boldsymbol{f}_{(j)}, \boldsymbol{\tau}_{(j)})p(\boldsymbol{\tau}_{(j)})}$}
    \RETURN $\tau_{(j)}\longleftarrow \tau^{'}_{(j)}, \boldsymbol{f}_{(j)}\longleftarrow \boldsymbol{f}^{'}_{(j)}$ \quad (Accept)
\ELSE
    \RETURN $\tau_{(j)}\longleftarrow \tau_{(j)}, \boldsymbol{f}_{(j)}\longleftarrow \boldsymbol{f}_{(j)}$ \quad (Keep)
\ENDIF
\end{algorithmic}
\end{algorithm}

After $\boldsymbol{f}_{(j)}$ is reparametrized, the posterior over the hyper-parameters for fixed $\boldsymbol{\eta}_{(j)}$ is expressed as follows:
\begin{align}
\begin{split}
p(\boldsymbol{\tau}_{(j)}|\boldsymbol{\eta}_{(j)},\mathcal{D},\boldsymbol{m})&\propto p(\mathcal{D},\boldsymbol{m}|\boldsymbol{f}_{(j)})p(\boldsymbol{\tau}_{(j)}) \\
&=p(\mathcal{D}|\boldsymbol{f}_{(j)})p(\boldsymbol{m}|\boldsymbol{f}_{(p+1)})p(\boldsymbol{\tau}_{(j)}),
\end{split}
\end{align}
in the binary values from $\mathcal{B}(1/1+\exp\{-(\lambda_0+f_i)\})$,
\begin{align}
\begin{split}
\displaystyle &p(\mathcal{D}|\boldsymbol{f}_{(j)}) \\
=&\prod^n_{i=1}\left[\left(\frac{1}{1+\exp(-(\lambda_0+f_i))}\right)^{x_i}\left(1-\frac{1}{1+\exp(-(\lambda_0+f_i))}\right)^{1-x_i} \right. \\
&\left. \left(\frac{1}{1+\exp(-(\lambda_0+f_i))}\right)^{y_{i1}}\left(1-\frac{1}{1+\exp(-(\lambda_0+f_i))}\right)^{1-y_{i1}}\right]^{m_i} \\
&\left[\left(\frac{1}{1+\exp(-(\lambda_0+f_i))}\right)^{x_i}\left(1-\frac{1}{1+\exp(-(\lambda_0+f_i))}\right)^{1-x_i} \right. \\
&\left. \left(\frac{1}{1+\exp(-(\lambda_0+f_i))}\right)^{y_{i2}}\left(1-\frac{1}{1+\exp(-(\lambda_0+f_i))}\right)^{1-y_{i2}}\right]^{1-m_i},
\end{split}
\end{align}
in the count values from $\mathcal{P}(\exp(\lambda_0+f_i))$,
\begin{align}
\begin{split}
\displaystyle &p(\mathcal{D}|\boldsymbol{f}_{(j)}) \\
=&\prod^n_{i=1}\left[\left(\frac{(\exp(\lambda_0+f_i))^{x_i}}{x_i !}\exp(-\exp(\lambda_0+f_i))\right)\left(\frac{(\exp(\lambda_0+f_i))^{y_{i1}}}{y_{i1} !}\exp(-\exp(\lambda_0+f_i))\right)\right]^{m_i} \\
&\left[\left(\frac{(\exp(\lambda_0+f_i))^{x_i}}{x_i !}\exp(-\exp(\lambda_0+f_i))\right)\left(\frac{(\exp(\lambda_0+f_i))^{y_{i2}}}{y_{i2} !}\exp(-\exp(\lambda_0+f_i))\right)\right]^{1-m_i},
\end{split}
\end{align}
or in the continuous values from $\mathcal{N}(\lambda_0+f_i,\sigma^2)$,
\begin{align}
\begin{split}
\displaystyle &p(\mathcal{D}|\boldsymbol{f}_{(j)}) \\
=&\prod^n_{i=1}\left[\left(\frac{1}{\sqrt{2\pi}\sigma}\exp\left\{-\left(\frac{x_i-(\lambda_0+f_i)}{2\sigma}\right)^{2}\right\}\right)\left(\frac{1}{\sqrt{2\pi}\sigma}\exp\left\{-\left(\frac{y_{i1}-(\lambda_0+f_i)}{2\sigma}\right)^{2}\right\}\right)\right]^{m_i} \\
=&\left[\left(\frac{1}{\sqrt{2\pi}\sigma}\exp\left\{-\left(\frac{x_i-(\lambda_0+f_i)}{2\sigma}\right)^{2}\right\}\right)\left(\frac{1}{\sqrt{2\pi}\sigma}\exp\left\{-\left(\frac{y_{i2}-(\lambda_0+f_i)}{2\sigma}\right)^{2}\right\}\right)\right]^{1-m_i}.
\end{split}
\end{align}
According to the probit model, (\ref{mi1}) and (\ref{mi2}),
\begin{align}
\begin{split}
\displaystyle &p(\boldsymbol{m}|\boldsymbol{f}_{(p+1)}) \\
=&\prod^n_{i=1}p(m_i=1|f_{(p+1)}(z_i))^{m_i}\left(1-p(m_i=0|f_{(p+1)}(z_i))\right)^{1-m_i}.
\end{split}
\end{align}

The prior parameters and missing observations are estimated using the RWMH algorithm.
In our generative models with the GP, the distribution of each parameter is as follows:
\begin{align}
\begin{split}
\boldsymbol{z}_i&\sim \mathcal{N}(\boldsymbol{0}_d, \boldsymbol{I}_d), \\
\tau_1&\sim \mathcal{N}(0, 1), \\
\tau_2&\sim \mathcal{IG}(5, 5), \\
\boldsymbol{\eta}_{(j)}&\sim \mathcal{N}(\boldsymbol{0}_n, \boldsymbol{I}_n), \\
\lambda_{0p^{\mathrm{out}}_j}&\sim \mathcal{N}(0, 1)\quad (\mathrm{if \ NMAR}), \\
\sigma&\sim \mathcal{N}(0, 1)\quad  (\mathrm{if \ Normal}).
\end{split}
\end{align}
In the sampling of latent variable $z_{ij}$, new candidate value $\hat{z}_{ij}$ is sampled from the proposal distribution, $\hat{z}_{ij}|\cdot \sim \mathcal{N}(z_{ij}, \upsilon^2)$, where $\upsilon^2$ is the variance of random walk, and it is adjusted to achieve an appropriate adoption rate.
The probability of adopting a new candidate value is $\min\left(1, \displaystyle \frac{p(\boldsymbol{y}|\hat{z}_{ij})p(\hat{z}_{ij})}{p(\boldsymbol{y}|z_{ij})p(z_{ij})}\right)$.
The same proposed distribution is used for the other parameters. However, for hyperparameter $\tau_2$, it is more efficient to sample from $\exp(\log(\hat{\tau}_2))|\cdot \sim \mathcal{N}(\exp(\log(\tau_2)), \upsilon^2)$ in a few cases.

The number of latent variables, $d$, is determined by comparing data combination models with different values of $d$ on the basis of the consistent Akaike information criterion (CAIC) statistic (Bozdogan, 1987).
\begin{align}
\mathrm{CAIC} = -2\ln p(\mathcal{D}|\mathcal{H})+\nu (\ln(n) + 1),
\end{align}
where $\nu$ is the effective number of parameters.

\section{Simulation study}
We generate simulation datasets and compare the prediction results obtained using the five methods listed in Table \ref{sim}. We compare the following methods: MM, LVM (EFAM)-MAR, LVM-NMAR, GPDCM-MAR, and GPDCM-NMAR. Among these, LVM-NMAR, GPDCM-MAR, and GPDCM-NMAR are the methods developed in this study. 
Although the true values of missing data are generally unknown in actual data analysis, we evaluate the prediction accuracy by creating multiple-source datasets based on the generated single-source dataset.

A single-source dataset with $200$ units and $12$ variables is generated. There are $10$ covariates, which consist of four continuous, three binary, and three count variables generated from the normal, Bernoulli, and Poisson distributions, respectively.
There are two outcome variables, which consist of one continuous and one binary variable.
For unit $i$, the GP canonical parameter is expressed as $\boldsymbol{\theta}_i=\boldsymbol{\lambda}_0+\boldsymbol{f}_i$. The missing indicator is defined as the sign of $u_i$, where $u_i=f_{(p+1)}(\boldsymbol{z}_i)+\varepsilon_i$, which divides the simulation dataset into two datasets.
Here, the GP latent function is generated by $\boldsymbol{z}_i\sim \mathcal{N}(\boldsymbol{0}_d, \boldsymbol{I}_d)$ and $\boldsymbol{f}_{(j)}\sim \mathcal{N}(\boldsymbol{0}, \boldsymbol{K}_{\boldsymbol{z}, \boldsymbol{\tau}})$.
The intercept is fixed as $\lambda_{0}=1$, and the dimension of the latent variables is fixed as two ($d=2$).

We evaluate the aforementioned methods using the mean squared error (MSE) between the true and predicted values.

Table \ref{sim} shows the mean and standard deviation of the MSE ratio obtained by applying each method to $300$ simulation datasets.
The MSE ratio for each method is standardized by the MSE ratio for MM; thus, the MSE ratio for MM is always one.
The MSE ratio of all the methods is $\mathrm{MSE} < 1.000$; thus, the results obtained using these methods are better than those obtained using MM.
The MSE ratio  of the GPDCM is smaller than that of the LVM in the MAR and NMAR cases, indicating that the introduction of the GP is quite effective  due to non-linearity of the data structure.
The MSE ratios of the methods that consider NMAR missingness are smaller than those of the methods that do not.
According to Table \ref{sim}, our method is useful for mixed observed data because the MSE ratio of the GPDCM for the NMAR case is the smallest.
\begin{table}[tbhp]
 \caption{MSE ratios obtained using five methods}
 \label{sim}
 \begin{center}
  \begin{tabular}{p{3.2cm}p{2.5cm}p{1.3cm}p{1.3cm}p{1.3cm}p{1.1cm}}
\noalign{\smallskip}\hline\noalign{\smallskip}
       Data combination &  Missing-data  &    \multicolumn{4}{l}{Outcome variables}       \\
\noalign{\smallskip}\cline{3-6}\noalign{\smallskip}
       method    & mechanism   &    \multicolumn{2}{l}{Continuous}    &  \multicolumn{2}{l}{Binary}       \\
            &     &   Mean   &  SD  &  Mean  &  SD    \\
\noalign{\smallskip}\hline\noalign{\smallskip}
       MM          & MAR   &   $1.000$  &  -        &   $1.000$  &  -   \\
       LVM (EFAM)  & MAR   &   $0.989$  &  $0.256$  &   $1.043$  &  $0.192$   \\
       LVM         & NMAR  &   $0.560$  &  $0.223$  &   $0.888$  &  $0.160$   \\
       GPDCM       & MAR   &   $0.283$  &  $0.223$  &   $0.727$  &  $0.187$   \\
       GPDCM       & NMAR  &   $0.126$  &  $0.120$  &   $0.530$  &  $0.179$   \\
\noalign{\smallskip}\hline\noalign{\smallskip}
  \end{tabular}
 \end{center}
\end{table}

\begin{figure}[tbhp]
  \vspace{-3mm}
  \begin{center}
    \includegraphics[scale=0.36]{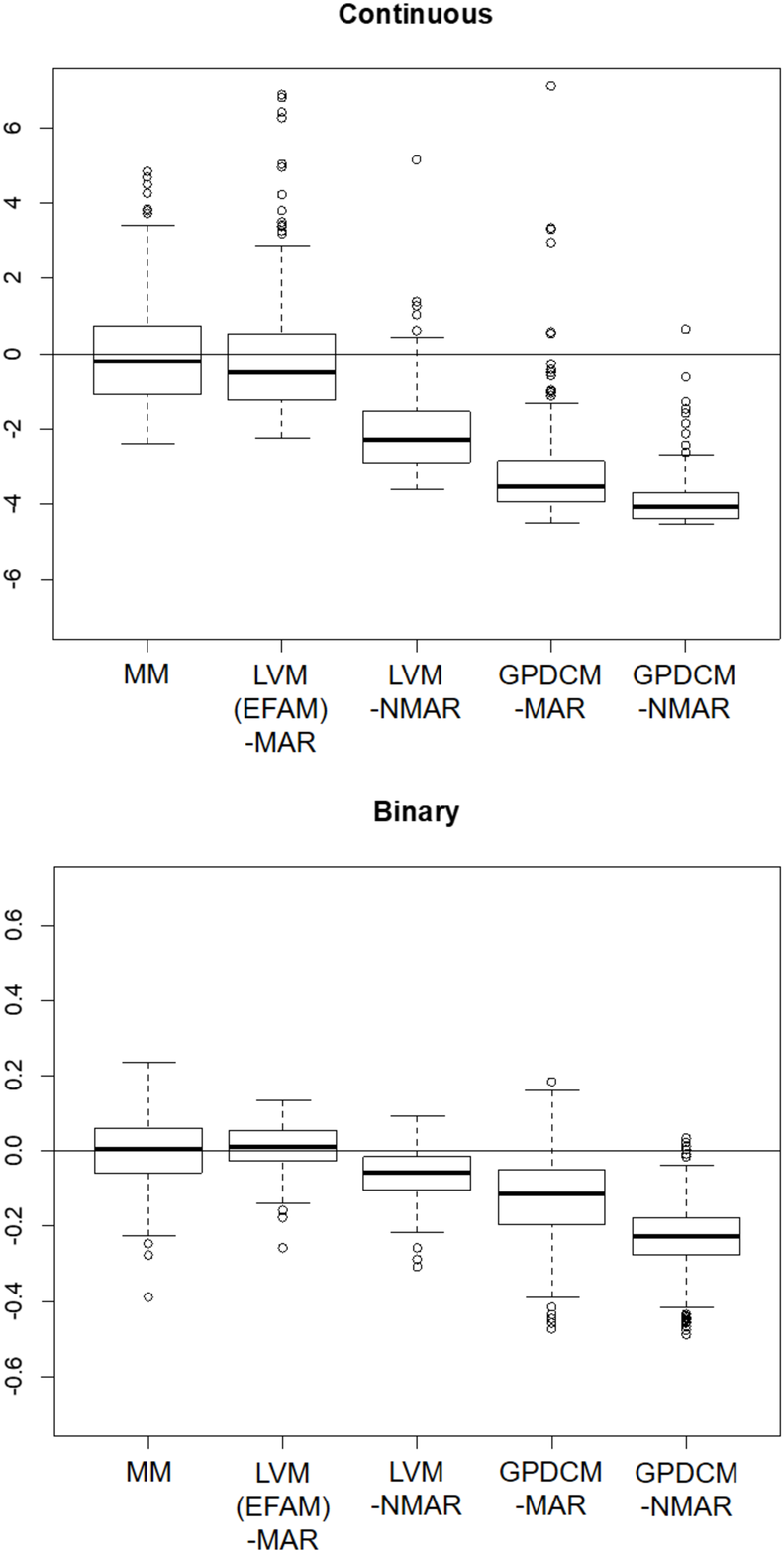}
  \end{center}
  \vspace{-2mm}
  \caption{Boxplot of MSE: prediction results of the continuous and binary outcome variables}
  \label{simbox}
\end{figure}

Figure \ref{simbox} shows the boxplot of the standardized MSE obtained from $300$ simulation datasets; the result agree with those discussed above.

\section{Illustrative application}
\begin{figure}[tbhp]
  \begin{center}
    \includegraphics[scale=0.37]{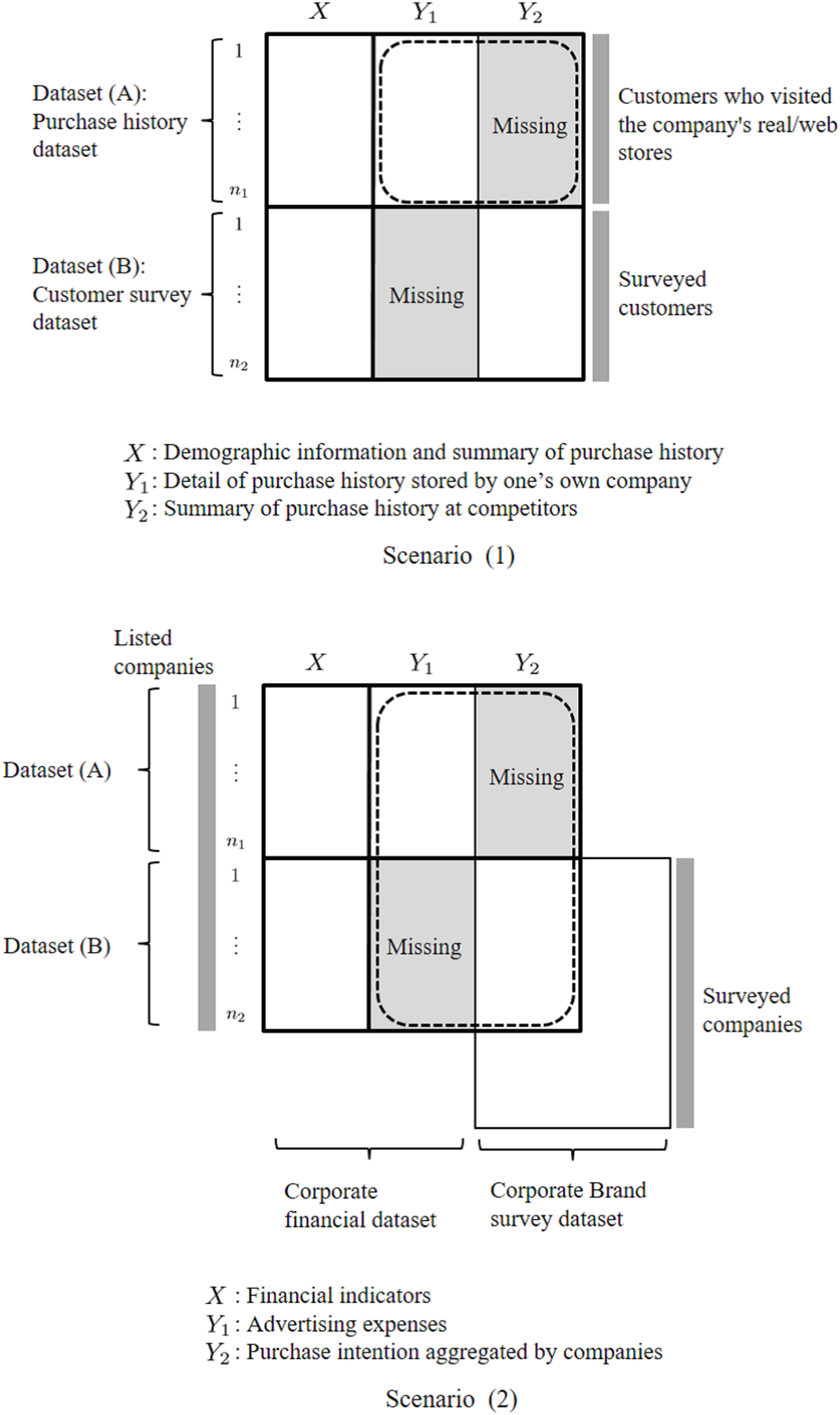}
  \end{center}
  \vspace{-3mm}
  \caption{Two scenarios of data combination}
  \label{scen}
\end{figure}

We describe the results of applying the proposed GPDCM-NMAR method to real-world data in two realistic scenarios where data combination techniques are required in marketing.
The two scenarios are shown in Figure \ref{scen}. The first is the data combination of a purchase history dataset and survey dataset for investigating the price elasticity of products, and the second is the data combination of a corporate financial dataset and survey dataset for analyzing the effectiveness of each company's advertising.
Typically, multiple-source datasets are obtained in both scenarios. However, in this work, we use valuable single-source datasets and divide them into multiple-source datasets to evaluate the methods in terms of the MSE between true and predicted values.

\subsection{Scenario (1): Data combination of purchase history dataset and survey dataset}
We use the consumer scan panel dataset provided by INTAGE Inc. This is a single-source dataset that includes the demographic information and purchase history for 2014.
We extract a subsample of consumers who purchase from a specific retailer and remove the purchase history for other retailers to mimic the ID-POS dataset.
In addition, we extract a random sample of the participants in the scan panel data and consider this as consumer survey data.
A retailer cannot investigate  the relationship between the purchase histories at his/her stores and competitors’ stores using only the purchase history data for his/her stores. However, a more rigorous analysis can be performed and more accurate predictions can be obtained by combining the survey data, including that of competitors, using data combination methods .

The dataset contains the $11$ variables, which are listed in Table \ref{sci}, for $300$ customers.
We divide the single-source dataset into two datasets, i.e., (A) purchase history data and (B) consumer survey data, with $n_1=n_2=150$, according to the probability obtained by the logistic function, $\delta(x_{i1})=1/(1+\exp({-x_{i1}}))$, of the ``Household income" variable.
As shown in Figure \ref{datadef}, these datasets assume the situations obtained from different units. The datasets consist of eight covariates, two observed outcome variables, and one missing outcome variable.
When both datasets are given, we apply the proposed method to deal with NMAR missing data.
\begin{table}[tbhp]
 \caption{Covariates and outcome variables in scenario (1)}
 \label{sci}
 \begin{center}
  \begin{tabular}{r@{\hspace{0.5cm}}p{8.2cm}p{1.8cm}}
\noalign{\smallskip}\hline\noalign{\smallskip}
     $j$  &  Items  &  Variables \\
\noalign{\smallskip}\hline\noalign{\smallskip}
     $1$  &  Demographic information: household income                      &    \\
     $2$  &  Demographic information: number of family members              &    \\
     $3$  &  Summary of purchase history at own stores: total purchase amount in January   &    \\
     $4$  &  Summary of purchase history at own stores: total purchase amount in February  &  \multicolumn{1}{c}{\multirow{2}{*}{$\boldsymbol{x}$}}  \\
     $5$  &  Summary of purchase history at own stores: total purchase amount in March     &    \\
     $6$  &  Summary of purchase history at own stores: total purchase amount in April     &    \\
     $7$  &  Summary of purchase history at own stores: total purchase amount in May       &    \\
     $8$  &  Summary of purchase history at own stores: total purchase amount in June      &    \\
          &                                                                 &    \\
     $9$  &  Detail of purchase history at own stores: total amount of products purchased in six months       &  \multicolumn{1}{c}{\multirow{2}{*}{$\boldsymbol{y}_1$}}  \\
    $10$  &  Detail of purchase history at own stores: quantity of products purchased in six months  &    \\
          &                                                                        &    \\ 
    $11$  &  Summary of purchase history at competitors stores: total purchase amount in June          &  \multicolumn{1}{c}{$\boldsymbol{y}_2$}  \\
\noalign{\smallskip}\hline\noalign{\smallskip}
  \end{tabular}
 \end{center}
\end{table}

The proposed methods estimate parameters and predict missing values using the MCMC algorithm.
We set $5,000$ iterations, $3,000$ burn-in samples, and five thinning intervals.

We reduce the number iterations by utilizing the estimated latent variable scores obtained from the GPLVM on the same dataset as the starting values.

The CAIC is used to select $d$.
We determine $d$ for the data combination methods other than MM by preparing candidates of one to four variables and comparing the CAIC.
Table \ref{sci_caic} shows the CAIC for each method standardized by the CAIC of one latent variable; $\mathrm{CAIC}=1$, where $d=1$.
We select $d$ with the lowest CAIC. Thus, we set $d=1$ for GPDCM-NMAR and GPDCM-MAR and $d=2$ for LVM(EFAM)-MAR and LVM-NMAR.
\begin{table}[tbhp]
 \caption{CAIC for the number of latent variables ($d$) in scenario (1)}
 \label{sci_caic}
 \begin{center}
  \begin{tabular}{p{3.4cm}p{2.7cm}p{2.1cm}p{2.1cm}p{2.1cm}p{2.1cm}}
\noalign{\smallskip}\hline\noalign{\smallskip}
       Data combination &  Missing-data  &    \multicolumn{4}{l}{CAIC ratio}       \\
\noalign{\smallskip}\cline{3-6}\noalign{\smallskip}
       method    & mechanism   &    \multicolumn{1}{c}{$d=1$}  &  \multicolumn{1}{c}{$2$}  &  \multicolumn{1}{c}{$3$}  &  \multicolumn{1}{c}{$4$}       \\
\noalign{\smallskip}\hline\noalign{\smallskip}
       LVM (EFAM) & MAR   &   \multicolumn{1}{r}{$1.000$}   &  \multicolumn{1}{r}{${\bf 0.988}$}  &   \multicolumn{1}{r}{$1.090$}  &   \multicolumn{1}{r}{$1.185$}     \\
       LVM        & NMAR  &   \multicolumn{1}{r}{$1.000$}   &  \multicolumn{1}{r}{${\bf 0.965}$}  &   \multicolumn{1}{r}{$1.091$}  &   \multicolumn{1}{r}{$1.214$}     \\
       GPDCM      & MAR   &   \multicolumn{1}{r}{${\bf 1.000}$}   &  \multicolumn{1}{r}{$1.006$}  &   \multicolumn{1}{r}{$1.110$}  &   \multicolumn{1}{r}{$1.131$}     \\
       GPDCM      & NMAR  &   \multicolumn{1}{r}{${\bf 1.000}$}   &  \multicolumn{1}{r}{$1.060$}  &   \multicolumn{1}{r}{$1.151$}  &   \multicolumn{1}{r}{$1.219$}     \\
\noalign{\smallskip}\hline\noalign{\smallskip}
  \end{tabular}
 \end{center}
\end{table}

Table \ref{sci_res} compares the MSE ratio for each method to that for MM. 
If the MSE ratio for a method is smaller than one, the prediction accuracy  of that method is better than that of MM.
As shown in Table \ref{sci_res}, the MSE ratios for all methods are smaller than one, and the MSE ratio for GPDCM-NMAR is the smallest.
\begin{table}[tbhp]
 \caption{MSE ratios for five methods in scenario (1)}
 \label{sci_res}
 \begin{center}
  \begin{tabular}{p{3.4cm}p{2.7cm}p{1.2cm}p{2.8cm}}
\noalign{\smallskip}\hline\noalign{\smallskip}
       Data combination &  Missing-data  &   \multirow{2}{*}{$d$}   &   \multirow{2}{*}{MSE ratio of $\boldsymbol{y}_2$}      \\
       method    & mechanism   &      &             \\
\noalign{\smallskip}\hline\noalign{\smallskip}
       MM         & MAR   &  -    &   $1.000$          \\
       LVM (EFAM) & MAR   &  $2$  &   $0.600$          \\
       LVM        & NMAR  &  $2$  &   $0.587$          \\
       GPDCM      & MAR   &  $1$  &   $0.063$          \\
       GPDCM      & NMAR  &  $1$  &   $0.049$          \\
\noalign{\smallskip}\hline\noalign{\smallskip}
  \end{tabular}
 \end{center}
\end{table}

\subsection{Scenario (2): Data combination of corporate financial dataset and survey dataset}
We use two datasets provided by Nikkei Research Inc., i.e., a corporate financial dataset and corporate brand survey dataset, for 2016.
The main variables in the corporate financial dataset are the items listed in the annual securities report.
This dataset contains missing values because a company may or may not publish its data on the basis of whether it is listed and its strategy.
The corporate brand survey dataset consists of the aggregated scores for the attitudes of each company obtained from Internet monitors.
Although this dataset provides information on consumer awareness and brand loyalty, which is not available from actual values such as financial variables, it contains missing values because it is difficult to survey all companies.
It is difficult to investigate the relationship between simultaneously obtained variables owing to numerous missing values. However, it is important to solve this problem in finance, and data combination methods are required for this purpose.

The two datasets are merged on the basis of the company units and treated as a single-source dataset.
This dataset contains $10$ variables, which are listed in Table \ref{zaimu_brand}, for $212$ Japanese companies. Only those companies whose advertising expenditures are publicly available and covered by the survey are selected.
We consider the total assets to be the company size and divide the single-source dataset into two datasets, (A) and (B), where $n_1=n_2=106$, according to the probability obtained by the logistic function, $\delta(x_{i1})=1/(1+\exp({-x_{i1}}))$.
These datasets assumes the situation shown in Figure \ref{datadef}, where there are $10$ covariates, one observed outcome variable, and one missing outcome variable.
Furthermore, it is assumed that datasets (A) and (B) are obtained from different units, as shown in Figure \ref{scen}.
When the datasets are given, we apply the proposed method to deal with NMAR missing data, where the MCMC settings are the same as those used in scenario (1).
\begin{table}[tbhp]
 \caption {Covariates and outcome variables in scenario (2)}
 \label{zaimu_brand}
 \begin{center}
  \begin{tabular}{r@{\hspace{0.5cm}}p{7.6cm}p{1.8cm}}
\noalign{\smallskip}\hline\noalign{\smallskip}
     $j$  &  Items     &  Variables \\
\noalign{\smallskip}\hline\noalign{\smallskip}
     $1$  &  Corporate finance: total assets             &    \\
     $2$  &  Corporate finance: net assets               &    \\
     $3$  &  Corporate finance: sales                    &    \\
     $4$  &  Corporate finance: operating profit margin  &  \multicolumn{1}{c}{\multirow{2}{*}{$\boldsymbol{x}$}}  \\
     $5$  &  Corporate finance: recurring margin         &    \\
     $6$  &  Corporate finance: net income margin        &    \\
     $7$  &  Corporate finance: ROA(Return On Assets)                      &    \\
     $8$  &  Corporate finance: ROE(Return On Equity)                      &    \\
          &                                              &    \\
     $9$  &  Corporate finance: advertising expenses     &  \multicolumn{1}{c}{$\boldsymbol{y}_1$}  \\
          &                                              &    \\ 
    $10$  &  Corporate brand survey: purchase intention  &  \multicolumn{1}{c}{$\boldsymbol{y}_2$}  \\
\noalign{\smallskip}\hline\noalign{\smallskip}
  \end{tabular}
 \end{center}
\end{table}
We determine $d$ by preparing candidates of one to four variables and comparing the CAIC.
Table \ref{fin_caic} shows the CAIC for each method standardized by the CAIC of one latent variable. We select $d$ with the lowest CAIC. Thus, $d=2$ for GPDCM-NMAR, and $d=4$ for LVM(EFAM)-MAR, LVM-NMAR, and GPDCM-MAR.
\begin{table}[tbhp]
 \caption{CAIC for the number of latent variables ($d$) in scenario (2)}
 \label{fin_caic}
 \begin{center}
  \begin{tabular}{p{3.4cm}p{2.7cm}p{2.1cm}p{2.1cm}p{2.1cm}p{2.1cm}}
\noalign{\smallskip}\hline\noalign{\smallskip}
       Data combination &  Missing-data  &    \multicolumn{4}{l}{CAIC ratio}       \\
\noalign{\smallskip}\cline{3-6}\noalign{\smallskip}
       method    & mechanism   &    \multicolumn{1}{c}{$d=1$}  &  \multicolumn{1}{c}{$2$}  &  \multicolumn{1}{c}{$3$}  &  \multicolumn{1}{c}{$4$}       \\
\noalign{\smallskip}\hline\noalign{\smallskip}
       LVM (EFAM) & MAR   &   \multicolumn{1}{r}{$1.000$}   &  \multicolumn{1}{r}{$0.182$}  &   \multicolumn{1}{r}{$0.130$}  &   \multicolumn{1}{r}{${\bf 0.112}$}     \\
       LVM        & NMAR  &   \multicolumn{1}{r}{$1.000$}   &  \multicolumn{1}{r}{$0.230$}  &   \multicolumn{1}{r}{$0.159$}  &   \multicolumn{1}{r}{${\bf 0.147}$}     \\
       GPDCM      & MAR   &   \multicolumn{1}{r}{$1.000$}   &  \multicolumn{1}{r}{$1.692$}  &   \multicolumn{1}{r}{$1.479$}  &   \multicolumn{1}{r}{${\bf 0.866}$}     \\
       GPDCM      & NMAR  &   \multicolumn{1}{r}{$1.000$}   &  \multicolumn{1}{r}{${\bf 0.869}$}  &   \multicolumn{1}{r}{$1.169$}  &   \multicolumn{1}{r}{$1.030$}     \\
\noalign{\smallskip}\hline\noalign{\smallskip}
  \end{tabular}
 \end{center}
\end{table}

Table \ref{fin_res} compares the MSE ratio for each data combination method. 
The MSE ratios for LVM-NMAR, GPDCM-MAR, and GPDCM-NMAR are smaller than that for MM.
Moreover, GPDCM-NMAR has the smallest MSE ratio and better prediction accuracy than MM.
\begin{table}[tbhp]
 \caption{MSE ratios for five methods in scenario (2)}
 \label{fin_res}
 \begin{center}
  \begin{tabular}{p{3.4cm}p{2.7cm}p{1.2cm}p{3.5cm}}
\noalign{\smallskip}\hline\noalign{\smallskip}
       Data combination &  Missing-data  &   \multirow{2}{*}{$d$}   &   \multirow{2}{*}{MSE ratio of ($\boldsymbol{y}_1, \boldsymbol{y}_2$)}      \\
       method    & mechanism   &      &             \\
\noalign{\smallskip}\hline\noalign{\smallskip}
       MM         & MAR   &  -    &   $1.000$          \\
       LVM (EFAM) & MAR   &  $4$  &   $0.815$          \\
       LVM        & NMAR  &  $4$  &   $0.755$          \\
       GPDCM      & MAR   &  $4$  &   $0.194$          \\
       GPDCM      & NMAR  &  $2$  &   $0.127$          \\
\noalign{\smallskip}\hline\noalign{\smallskip}
  \end{tabular}
 \end{center}
\end{table}

\begin{figure}[tbhp]
  \vspace{-3mm}
  \begin{center}
    \includegraphics[scale=0.55]{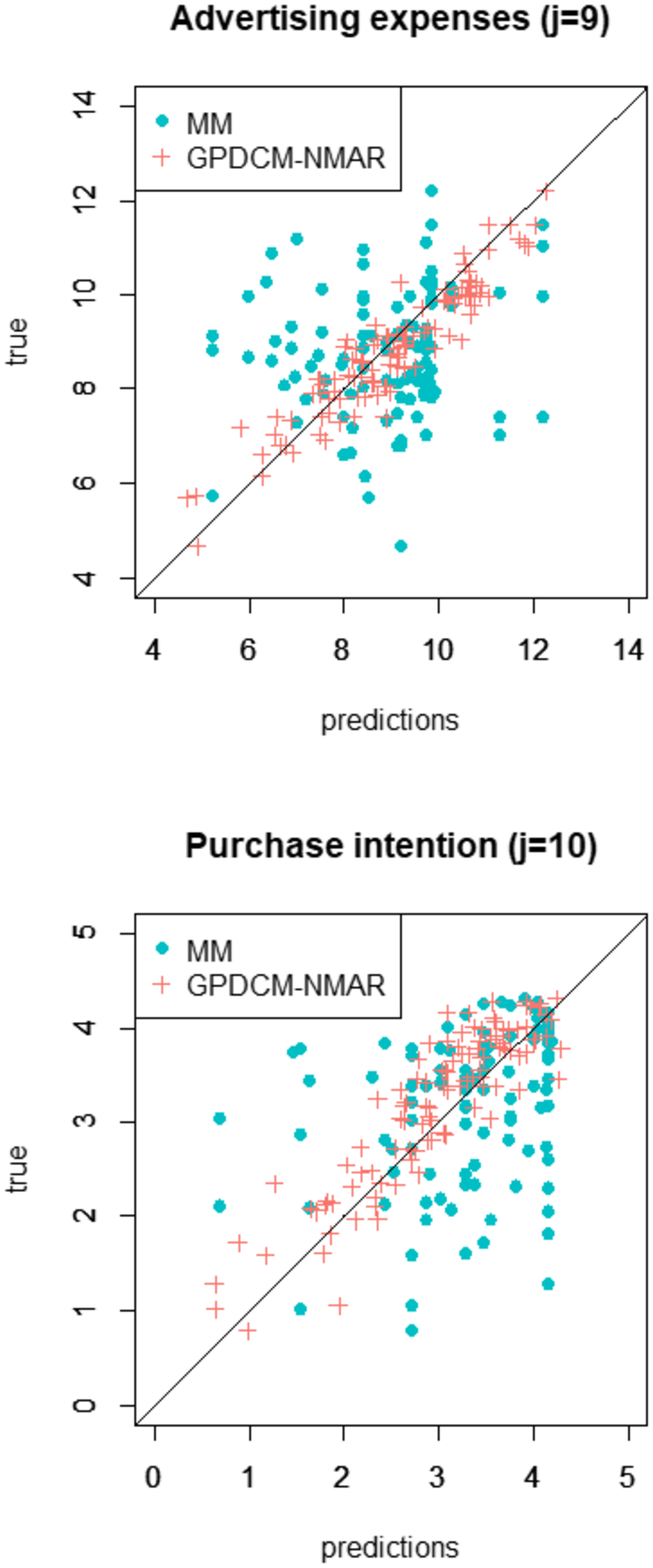}
  \end{center}
  \vspace{-2mm}
  \caption{Prediction results of two outcome variables in scenario (2)}
  \label{fin_plot}
\end{figure}

The plots of the true and predicted values obtained using GPDCM-NMAR are shown in Figure \ref{fin_plot}.
In both plots, MM shows poor prediction accuracy. The prediction accuracy is considerably improved by imputing observed values under the assumption of NMAR missing data.
These plots show that GPDCM-NMAR is useful for the data combination of multiple-source datasets with selection bias.

\begin{figure}[tbhp]
  \vspace{-3mm}
  \begin{center}
    \includegraphics[scale=0.52]{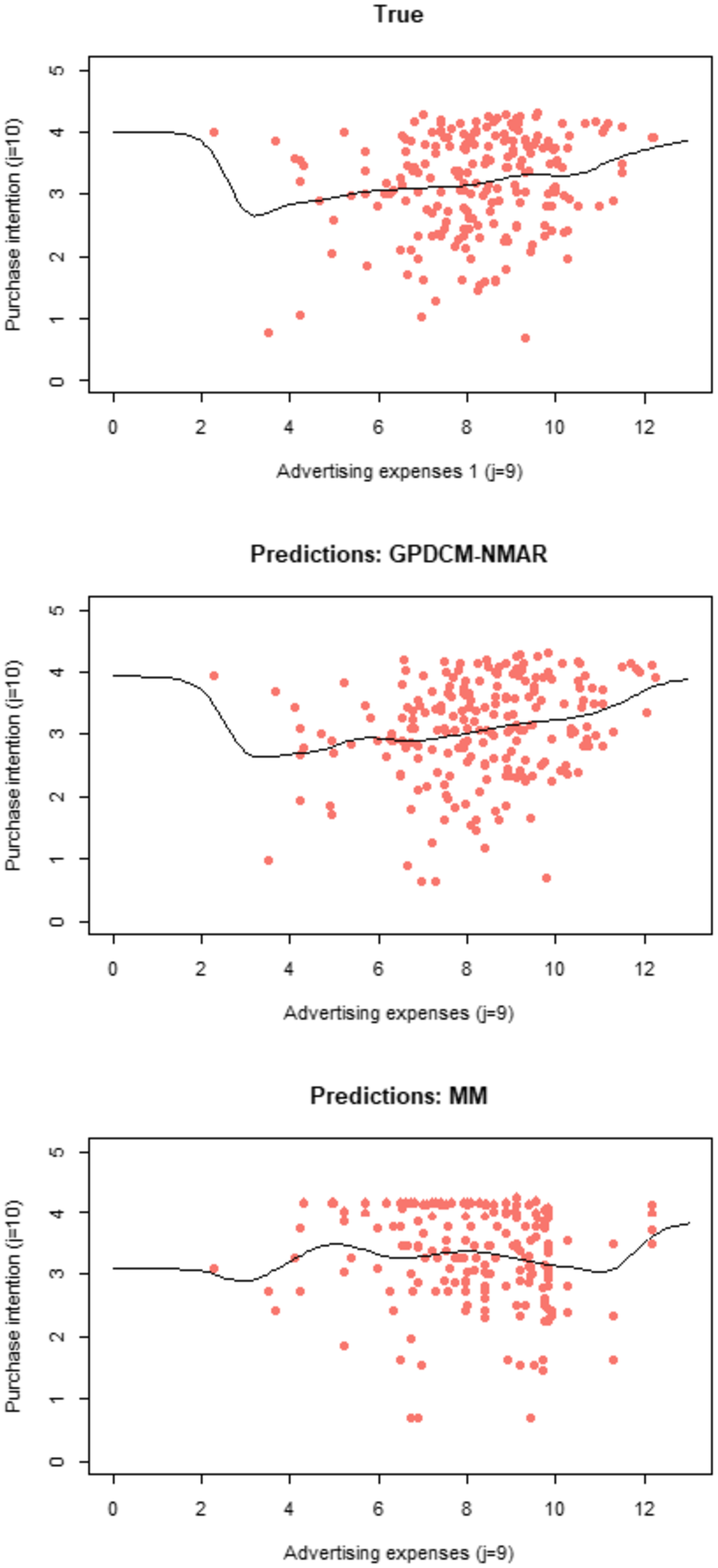}
  \end{center}
  \vspace{-2mm}
  \caption{Relationship between outcome variables in scenario (2)}
  \label{fin_plot2}
\end{figure}
Figure \ref{fin_plot2} shows the scatter plots between the advertising expense and purchase intention obtained from the true single-source data and the results predicted by GPDCM-NMAR and MM, where the solid line is the kernel regression function with the bandwidth determined using Silverman's rule of thumb.
The scatter plot and regression function obtained using GPDCM-NMAR are quite similar to those obtained from the true dataset, whereas those obtained using MM are considerably different.

\section{Conclusions}
We propose a GPDCM under NMAR missingness to integrate the multiple-source datasets of mixed observed variables into a single-source dataset.
The proposed method can capture the nonlinear relationship between observed and latent variables by modeling observed variables and a missing indicator using GP latent functions.
As observational data may experience selection bias, this method can handle NMAR and MAR missing data by assuming that the missing indicator depends on latent variables.
The proposed method is compared with other methods using a simulation study and real-world data analysis assuming NMAR missingness. Results show that the MSE ratio for our method is the smallest, and the prediction accuracy is improved.

Our method can be applied to mixed discrete and continuous variables by modeling a distribution in the exponential family using GP latent functions.
In the example described in the introduction, there are multiple-source datasets of the purchase history and web-ad exposure. The covariates of these datasets are typically demographic variables such as gender, age, occupation, and residence area.
These variables are qualitative, and they are frequently used as binary variables converted into dummy variables.
In such situations, the application of our method to mixed observed data can be useful .

A limitation of our method is that its calculation cost is high because it estimates parameters and predicts missing values using the MCMC algorithm.
Scalability can be improved by utilizing approximation methods such as variational Bayesian inference and sparse kernel estimation.

%\begin{acknowledgement}
%The authors would like to thank reviewers and editors for their very helpful comments on
%this manuscript.
%\end{acknowledgement}




\end{document}